\documentclass{jfm}
\usepackage{graphicx}
\usepackage{epsfig,xcolor}

\shorttitle{Universal fluctuations in the bulk of Rayleigh-B\'enard turbulence}
\shortauthor{Y.-C. Xie, B.-Y.-C. Cheng, Y.-B. Hu and K.-Q. Xia}
\title{Universal fluctuations in the bulk of Rayleigh-B\'enard turbulence}
\author{Yi-Chao Xie \aff{1,2}
  \corresp{\email{ycxie@cuhk.edu.hk}},
  Bu-Ying-Chao Cheng\aff{1},
 Yun-Bing Hu\aff{1,2},
 \and Ke-Qing Xia\aff{2,1} \corresp{\email{xiakq@sustech.edu.cn}}}
\affiliation{\aff{1}Department of Physics, The Chinese University of Hong Kong, Shatin, Hong Kong, China
\aff{2}Center for Complex Flows and Soft Matter Research \& Department of Mechanics and Aerospace Engineering, Southern University of Science and Technology, Shenzhen, 518055, China}
\begin{document}
\maketitle

\begin{abstract}

We present an investigation of the root-mean-square (rms) temperature $\sigma_T$ and the rms velocity $\sigma_w$ in the bulk of Rayleigh-B\'enard turbulence, using new experimental data from the current study and experimental and numerical data from previous studies. We find that, once scaled by the convective temperature $\theta_*$, the value of $\sigma_T$ at the cell centre is a constant, i.e. $\sigma_{T,c}/\theta_* \approx 0.85$, over a wide range of the Rayleigh number ($10^{8}\leq Ra\leq 10^{15}$) and the Prandtl number ($0.7\leq Pr \leq 23.34$), and is independent of the surface topographies of the top and bottom plates of the convection cell. A constant close to unity suggests that $\theta_*$ is a proper measure of the temperature fluctuation in the core region. On the other hand, $\sigma_{w,c}/w_*$, the vertical rms velocity at the cell centre scaled by the convective velocity $w_*$, shows a weak $Ra$-dependence ($\sim Ra^{0.07\pm0.02}$) over $10^8\leq Ra\leq 10^{10}$ at $Pr\sim4.3$ and is independent of plate topography. Similar to a previous finding  by He \& Xia ({\it Phys. Rev. Lett.,} vol. 122, 2019, 014503), we find that the rms temperature profile $\sigma_T(z)/\theta_*$ in the region of the mixing zone with a mean horizontal shear exhibits a power-law dependence on the distance $z$ from the plate, but now the universal profile applies to both smooth and rough surface topographies and over a wider range of $Ra$. The vertical rms velocity profile $\sigma_w(z)/w_*$ obey a logarithmic dependence on $z$. The study thus demonstrates that the typical scales for the temperature and the velocity are the convective temperature $\theta_*$ and the the convective velocity $w_*$, respectively. Finally, we note that $\theta_*$ may be utilised to study the flow regime transitions in the ultra-high-$Ra$-number turbulent convection.

\end{abstract}

\section{Introduction}\label{sec_intro}

Rayleigh-B\'enard convection (RBC), a fluid layer confined between two plates heated from below and cooled from above, continues to attract attention not only because of its relevance to geophysical and astrophysical flows, but also owing to the fact that it's an idealised model for the study of thermally driven turbulence (for reviews, see, for example, \citet* {Ahlers2009RMP,Xia2010ARFM,Chilla2012EPJE,Xia2013TAML}). In RBC, buoyancy injects energy into the turbulence and creates  vigorous velocity and temperature fluctuations. Understanding the dynamics of these fluctuations has been one of the central issues  \citep{GL2004PoF}. The RBC system is controlled by two dimensionless parameters, i.e. the Rayleigh number $Ra=\alpha g \Delta T H^3/(\kappa\nu)$ and the Prandtl number $Pr=\nu/\kappa$, where $g$ is the gravitational acceleration constant, $H$ the height of the convection cell, $\Delta T$ the temperature difference across the top and bottom plates, $\alpha, \kappa$ and $\nu$ are the thermal expansion coefficient, the thermal diffusivity and the kinematic viscosity of the working fluid, respectively. 

Two turbulence states with different probability density functions (PDFs) of the temperature at the cell centre were observed in turbulent RBC, i.e. a `soft turbulence' state with a Gaussian PDF and a `hard turbulence'  state with an exponential PDF \citep*{Heslot1987PRA}. In the latter, the temperature fluctuation $\sigma_T$ normalised by $\Delta T$ obeys a $-0.145$ power law with $Ra$. A mixing-length model was proposed to explain this dependence \citep{Castaing1989JFM}. However, later studies show that $\sigma_T/\Delta T$ not only depends on $Pr$ but also on the plate topography even for the most widely studied aspect-ratio-unity cylindrical cells (see table \ref{Tab1} for a summary). It is thus natural to ask is there any universal behaviours of the temperature fluctuations in the bulk flow. 

In addition to the value of $\sigma_T$ at the cell centre, the functional form of the root-mean-square (rms) temperature profile $\sigma_T(z)$ is also of great interest as its shape determines the transport properties across the boundary. Based on different assumptions of the local force balance, theory predicts different profiles of $\sigma_T(z)$ \citep{Adrian1996}. Only very recently, a clear understanding of $\sigma_T(z)$ is obtained \citep{He2019PRL}: While in a region with mean horizontal shear (the viscous force balances the inertia force), $\sigma_T(z)$ obeys a power-law dependence on $z$; in regions with abundant plume emissions (the buoyancy balances the inertia force), $\sigma_T(z)$ is in a logarithmic form, which are true for idealised case, i.e. turbulent RBC in cells with smooth surfaces (``smooth cell"). With the presence of roughness on the top and bottom boundaries (``rough cell"), $\sigma_T$ enhances considerably \citep{Du2001PRE}. An interesting question is will $\sigma_T$ show universal behaviours in cells with different plate topographies? In addition to the shape of $\sigma_T(z)$, there is no generally accepted characteristic temperature scale in the bulk. For instance, $\Delta T$ has been widely used as a typical temperature scale, but the maximum value $\sigma_T(z){_{max}}$ is also used sometimes (see, for example, \citet{Wang2018JFM}). The $Ra$-dependence of $\sigma_T(z)$ scaled by different scales exhibit different features: While $\sigma_T(z)/\Delta T$ for different $Ra$ differs (\citet*{Sun2008JFM}; \citet{Ahlers2012PRL}), $\sigma_T(z)/\sigma_T(z){_{max}}$ for different $Ra$ collapses better in the mixing zone. Therefore, the typical temperature scale in the bulk is not clear at present.

Compared to numerous studies on temperature fluctuations, the investigation on velocity fluctuations in turbulent RBC is scarce. The velocity fluctuation is usually studied in terms of a Reynolds number based on the vertical rms velocity $\sigma_{w}$, i.e. $Re_{\sigma_w}=\sigma_wH/\nu$. It is found that $Re_{\sigma_w}$ scales with $Ra$ to a 0.5 power law, consistent with the free-fall like argument (\citet*{Shen1995PRL}; \citet{Daya2001PRL}; \citet{Qiu2004PoF}; \citet*{Shang2008PRL}). To the best of our knowledge, there is almost no direct measurement of velocity fluctuations in rough cells. The typical velocity scale in the bulk flow also remains to be explored.  

In this paper, we present an investigation of the temperature and the velocity fluctuations in the bulk of turbulent RBC. We demonstrate that the typical temperature and velocity scales in the bulk are, respectively, the convective temperature $\theta_*$ and the convective velocity $w_*$. Once scaled by these quantities, the fluctuations in the bulk exhibit universal behaviours. These findings shed new light on the bulk dynamics in convective turbulence. 

\section{The experimental setup and relevant parameters}\label{sec_exp}

 \begin{table}
  \begin{center}
\def~{\hphantom{0}}
  \begin{tabular}{lcccccr}
  Ref. &        $Ra$                                             & $Pr$ &  $A$  & $\gamma$ & Geometry & Surface type\\
     $(a)$   & $1.2\times10^8 \sim 6.5\times10^{10}$ & 0.7 & 0.36  &-0.147  & Cylinder   &S\\
     $(b)$   & $7.7\times10^7 \sim1.0\times10^{15}$ & 0.7 & 0.37   &-0.145 & Cylinder    &S\\
     $(c)$   & $4.8\times10^8 \sim 5.8\times10^9$   & 5.4 & 0.192  & -0.14& Cylinder     &R\\
     $(d)$   & $2.5\times10^8 \sim 3.9\times10^9$   & 5.46 & N.A.  &-0.10 &    Cylinder& S \\
      $(d)$   & $2.5\times10^8 \sim 3.9\times10^9$   & 5.46 & N.A.  &-0.48 & Cube& S \\
     $(e)$   & $5.7\times10^8 \sim 1.1\times10^{10}$   & 7 & 5.9 &-0.35& Cube & S \\
     $(f)$   & $1.0\times10^7 \sim 2.0\times10^9$   & 5.2 & N.A.  &-0.18& Cylinder & S \\
     $(g)$   & $3.6\times10^8 \sim 7.6\times10^9$    &  4.3 & $0.066\sim0.10$ &-0.1 &Cylinder & S \& R\\
     $(h)$    & $3.0\times10^{7} \sim 3.0\times10^{9}$ &  4.3& 9.38&-0.35& Cube &S\\
     $(i)$    & $1.0\times10^{11} \sim 4.2\times10^{11}$ &  12.3 & N.A. &-0.17 & Cylinder &S\\
     $(j)$   & $4.0\times10^9 \sim 1.3\times10^{11} $  & 23.34 &$0.08$ & $ -0.09$ & Cylinder &R\\          
       present & $1.6\times10^8 \sim 8.8\times10^9$ & 4.3 & $0.001 \sim 0.12$&$-0.10 \sim 0.12$ &Cylinder&S \& R\\
  \end{tabular}
  \caption{Scaling of the temperature fluctuations ($\sigma_{T,c}/\Delta T=ARa^{\gamma}$) at the cell centre.  In the surface type column, ``S" stands for ``smooth surface cell" and ``R" stands for ``rough surface cell". The references are: $(a)$ \cite{Castaing1989JFM}, $(b)$ \cite{Niemela2000Nat}, $(c)$ \cite{Du2001PRE}, $(d)$ \cite{Daya2001PRL}, $(e)$ \cite{Wang2003EPJB}, $(f)$ \cite{Lakkaraju2012PRE}, $(g)$ \cite{Wei2014JFM}, $(h)$ \citet{Kaczorowski2014JFM}, $(i)$ \cite{Wei2016JFM}; $(j)$ \cite{XieXia2017JFM}.}
  \label{Tab1}
  \end{center}
\end{table}

The experiment was carried out in an upright cylindrical cell with an aspect ratio $\Gamma=D/H\approx 1$, where $D=192$ mm is the cell diameter and $H= 202$ mm is its height. The design and construction of the cell can be found in \citet{XieXia2017JFM}. We mention only its essential features here. The cell consists of three parts, i.e. a copper top plate, a copper bottom plate and a Plexiglas sidewall. The bottom plate was heated by rubber electrical heaters and the top plate was cooled by passing temperature controlled water through a chamber fitted onto its top surface. The temperature of the top (bottom) plate was measured using four (five) thermistors from which we calculated $Ra$, $Pr$ and the Nusselt number $Nu=QH/(\chi\Delta T)$, where $Q$ is the heat flux supplied at the bottom plate and $\chi$ the thermal conductivity of the fluid. Deionzed and degassed water was used as the working fluid with a mean fluid temperature kept at a constant of $40$ $^oC$. Thus the Prandtl number was $Pr =4.34$. By changing $\Delta T$, we achieved a $Ra$ range of $1.59\times 10^8\leq Ra\leq 8.82\times10^9$. To study the effects of wall roughness on the fluctuations of the temperature and the velocity, another set of measurements in a rough cell were made. The roughness elements were in the form of square-latticed pyramids with a height $h$ of 8 mm. The heat transport in rough cells show three regimes depending on $Ra$ \citep{XieXia2017JFM}. The $Ra$ range in the present study in the rough cell is in the heat-transport-enhanced regimes.

The temperature fluctuation, $\sigma_T=\sqrt{\langle(T-\langle T\rangle)^2\rangle}$, inside the cell was measured using a waterproof thermistor with a head diameter of $0.3$ mm and a response time of $30$ ms, where $\langle \cdots\rangle$ stands for time averaging. Two sets of temperature fluctuation measurements were made. In the first set, the temperature fluctuations at the centre $\sigma_{T,c}$ of the smooth cell and the rough cell were measured as a function of $Ra$. In the second set, the vertical rms temperature profiles $\sigma_T(z)$ were measured in the rough cell. The location of $z=0$ mm was set at the valley of the roughness elements on the bottom plate. At each vertical position $z$, temperature time trace was measured for one-hour with a sampling rate of 15 Hz. Good care was taken to ensure that the system was isolated from the surrounding temperature variations. For the detailed thermal isolation method, we refer to  \citet*{Xie2018PRL}.

The vertical velocity fluctuation, characterised by a Reynolds number $Re_{\sigma_w}=\sigma_wH/\nu$, was measured at the centre of the rough cell, using a laser Doppler velocimeter (LDV). Here $\sigma_w=\sqrt{\langle(w-\langle w\rangle)^2\rangle}$ is the vertical rms velocity. The flow was seeded with tracer particles with a diameter of 2.893 $\mu$m. The LDV sampling rate was $\sim$ 30 Hz. Typical measurement lasted for 12 hours to obtain sufficient statistics of the second-order quantity like the rms velocity. 

Consider the region outside the boundary layers, the relevant physical parameters governing the flow dynamics are $\alpha g$, $Q_0$ and $H$, where $Q_0=Q/(\rho c_p)$ is the specific heat flux, $\rho$ and $c_p$ are, respectively, the density and the specific heat capacity of the working fluid. A simple dimensional analysis yields the convective temperature  $\theta_*$ and the convective velocity $w_*$: $\theta_* \equiv Q_0^{2/3}/(\alpha g H)^{1/3}$ and $w_*\equiv(\alpha g H Q_0)^{1/3}$  \citep{Deardorff1970}. The two scales can be expressed in dimensionless forms in terms of $Ra$, $Pr$ and $Nu$:
\begin{equation}
\theta_*/\Delta T=Nu^{2/3}/(RaPr)^{1/3} \qquad Re_{w_*}=w_*H/\nu=(RaNuPr^{-2})^{1/3}
\label{equ1}
\end{equation}

\section{Results and discussions}\label{sec_results}

\subsection{The temperature fluctuation in the bulk}
 
\begin{figure}
      \center
     \includegraphics[width=\textwidth]{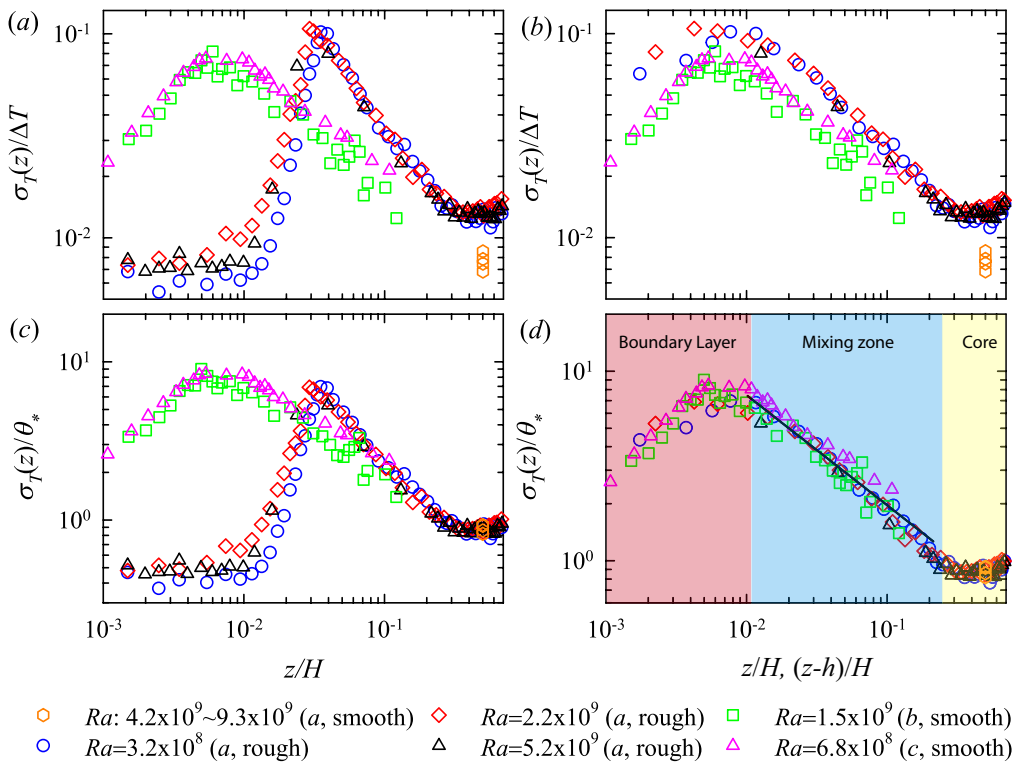}
      \caption{Measured rms temperature profiles  $\sigma_T(z)$ along the cell centreline. The vertical axis is scaled by $\Delta T$ (upper panel) and by $\theta_*$  (lower panel). The horizontal axes are scaled by the cell height $H$.  The legends with `smooth' and `rough' mean smooth cells and rough cells, respectively. In $(b,d)$, the distance $z$ for the data measured in the rough cell is offset by the roughness height $h$. The solid line in ($d$) is a power law fit to the data in the range $10^{-2}\leq z/H\leq 2\times10^{-1}$, i.e. $\sigma_T/\theta_*=0.53\times (z/H)^{-0.57\pm0.03}$. The data sources are $a$: present, $b$: \cite{Du2000JFM} and $c$: \cite{Wei2014JFM}. }
      \label{fig1}
    \end{figure}

We first study the rms temperature profiles. Figures \ref{fig1}($a,c$) show $\sigma_T(z)$ normalised, respectively, by the temperature difference across the top and bottom plates $\Delta T$ and the convective temperature $\theta_*$ measured in the rough cell for $Ra=3.2\times10^8, 2.2\times10^9$ and $5.2\times10^9$. The rms temperatures ($\sigma_{T,c}$) measured at the centre of the smooth cell for $8.9\times10^8\leq Ra\leq 9.3\times10^9$ are plotted as hexagons. For comparison, we also plot in these figures $\sigma_T(z)$ measured in smooth cells from \cite{Du2000JFM} at $Ra=1.5\times 10^9$ and from \cite{Wei2014JFM} at $Ra=6.8\times10^8$. 
The horizontal axes are normalised by $H$. 

It is seen that the profiles measured in either a smooth cell or a rough cell for different $Ra$ collapse onto each other outside the thermal boundary layer (TBL). With increasing $z$, $\sigma_T(z)$ increases to a maximum and then decreases gradually when moving towards the cell centre. These observations are in general consistent with earlier studies \citep{Lui1998PRE}. However, considerable differences between $\sigma_T(z)/\Delta T$ measured in smooth cells and in rough cells are observed (figure \ref{fig1}$a$). First, the outer edge of the TBL, i.e. the peak position of $\sigma_T(z)/\Delta T$, is shifted towards the cell centre, which is because the motionless fluids trapped in the bottom of valleys between roughness elements have very small temperature fluctuations. Second, $\sigma_T$ enhances considerably near the edge of the TBL in rough cells owing to more thermal plumes being emitted from the tip of roughness elements. This enhancement is also true at the cell centre since the data in the smooth cell are systematically smaller than those in the rough cell. However, we note, even for a smooth cell, \citet{Ahlers2012PRL} showed that $\sigma_T(z)/\Delta T$ for different $Ra$ does not collapse in the classical regime of turbulent RBC. The above results suggest that $\Delta T$ is not the suitable characteristic scale for $\sigma_T$. When scaled with $\theta_*$, the temperature fluctuations at the cell centre, for both the smooth and the rough cells, collapse onto each other (but not for the profile in the mixing zone), and the magnitude of the fluctuation at the edge of the TBL also becomes comparable in two cases (figure \ref{fig1}$c$). To compare directly the rms temperature profiles in the bulk region, in figures \ref{fig1}$(b,d)$, the distance $z$ of the data measured in the rough cells is offset by the roughness height $h$. Remarkably, compared with $\sigma_T(z)/\Delta T$ (figure \ref{fig1}$b$), $\sigma_T(z)/\theta_*$ in smooth and rough cells collapses onto each other outside the TBL (figure \ref{fig1}$d$), suggesting that $\theta_*$ is a characteristic temperature scale in the bulk of turbulent convection.

  \begin{figure}
      \center
     \includegraphics[width=\textwidth]{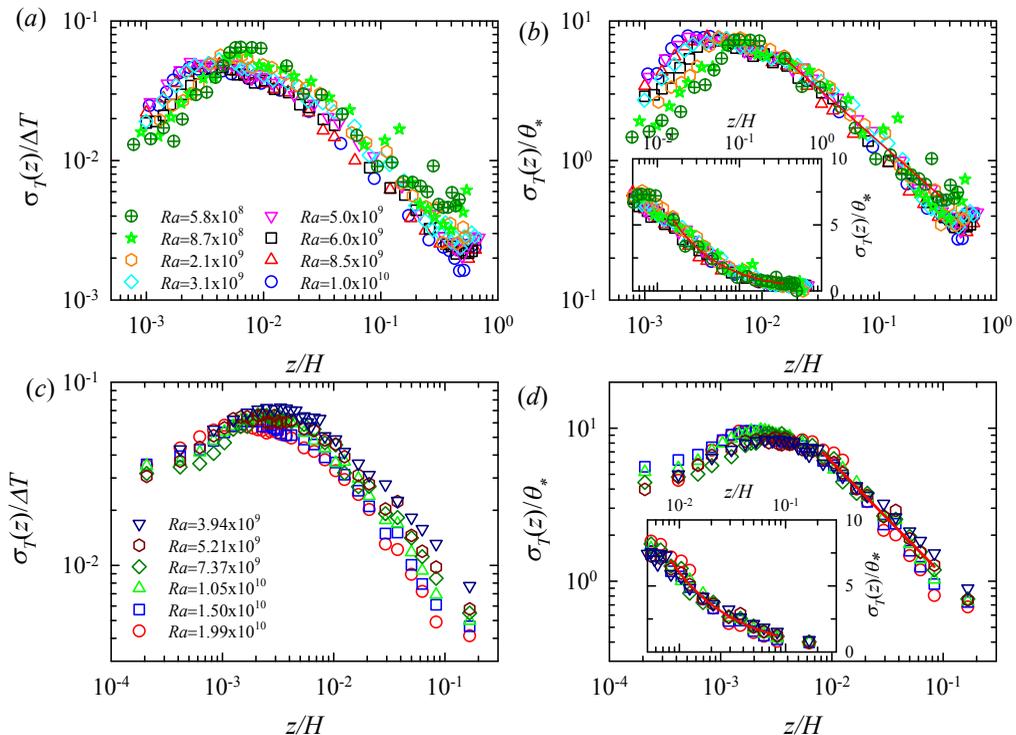}
      \caption{Measured rms temperature profiles along the centreline in a cubic cell ($a,b$)  and in a rectangular cell ($c,d$). The vertical axe are normalised by $\Delta T$ ($a,c$) and by $\theta_*$ ($b,d$). The solid lines in ($b,d$) are power-law fits, i.e.  ($b$) $\sigma_T(z)/\theta_*=0.25(z/H)^{-0.74\pm0.02}$ and ($d$) $\sigma_T(z)/\theta_*=0.20(z/H)^{-0.74\pm0.02}$. Insets of ($b,d$): $\sigma_T(z)/\theta_*$ in linear-log plots. The data in the cubic cell are taken from \cite{Wang2003EPJB} and those in the rectangular cell from \cite{Sun2008JFM}. }
      \label{fig2}
      \end{figure}     
     
We now examine the universal profiles shown in figure \ref{fig1}($d$) in detail. They can be divided into three regions, i.e. a TBL region, a mixing zone, and a core region, which are marked with different background colours. In the mixing zone spanning $3\times 10^{-2}\leq z/H\leq 2.5\times10^{-1}$, a power law $\sigma_T/\theta_*=0.53\times (z/H)^{-0.57\pm0.03}$ fits the data nicely. This scaling exponent is consistent with results in literature \citep{Wei2016JFM,Wang2018JFM,He2019PRL}, and also close to the theoretical prediction of $-1/2$ by \citet{Adrian1996}. This observation further supports that $\sigma_T(z)$ obeys a power-law decay in a region where a mean horizontal shear exists, and it is true for both smooth cells and rough cells. In the core region ($0.25\leq z/H\leq 0.65$), $\sigma_T(z)/\theta_*$ is nearly a constant independent of $Ra$ and the plate topography, suggesting a universal temperature fluctuation here. 
            
We next test the convective temperature in other cell geometries. Figures \ref{fig2}($a,b$) show $\sigma_T(z)$ normalised respectively by $\Delta T$ and $\theta_*$ measured along the centreline in a cubic cell. The rms temperature profiles obtained along the cell centreline in a rectangular cell are shown in figures \ref{fig2}$(c,d)$. The data in the cubic cell are taken from \citet{Wang2003EPJB} and those in the rectangular cell are from \citet{Sun2008JFM}. Let's focus on the cubic case first. It is seen that the originally $Ra$-dependent profiles (figure \ref{fig2}$a$) collapse onto each other once they are scaled by $\theta_*$ (figure \ref{fig2}$b$). The solid line in figure \ref{fig2}($b$) is a power law fit to the data in the range $2\times 10^{-2}\leq z/H\leq 3\times10^{-1}$, yielding $\sigma_T(z)/\theta_*=0.25(z/H)^{-0.74\pm0.02}$. Note that $\sigma_T(z)/\theta_*$ for different $Ra$ do not collapse inside the TBL. This is because $\theta_*$ is a temperature scale for the bulk, so is not applicable in the near wall region where the viscosity and the thermal diffusivity dominate the dynamics. Similar behaviours are observed in the rectangular cell, i.e. when compared with $\sigma_T(z)/\Delta T$, the $\sigma_T(z)/\theta_*$ collapses better for different $Ra$, and it can be fitted by a power-law with a scaling exponent of $-0.74\pm0.02$ in the mixing zone. The insets of figures \ref{fig2}($b,d$) plot the same data as the main figures but in log-linear plots. They clearly show that these profiles cannot be fitted by logarithmic functions. Note that the scaling exponent is different from the cylindrical cells, suggesting that turbulent fluctuations in the bulk are cell-geometry-dependent, which may be partially attributed to the geometry-dependence of the large-scale flow dynamics.

      \begin{figure}
      \center
     \includegraphics[width=\textwidth]{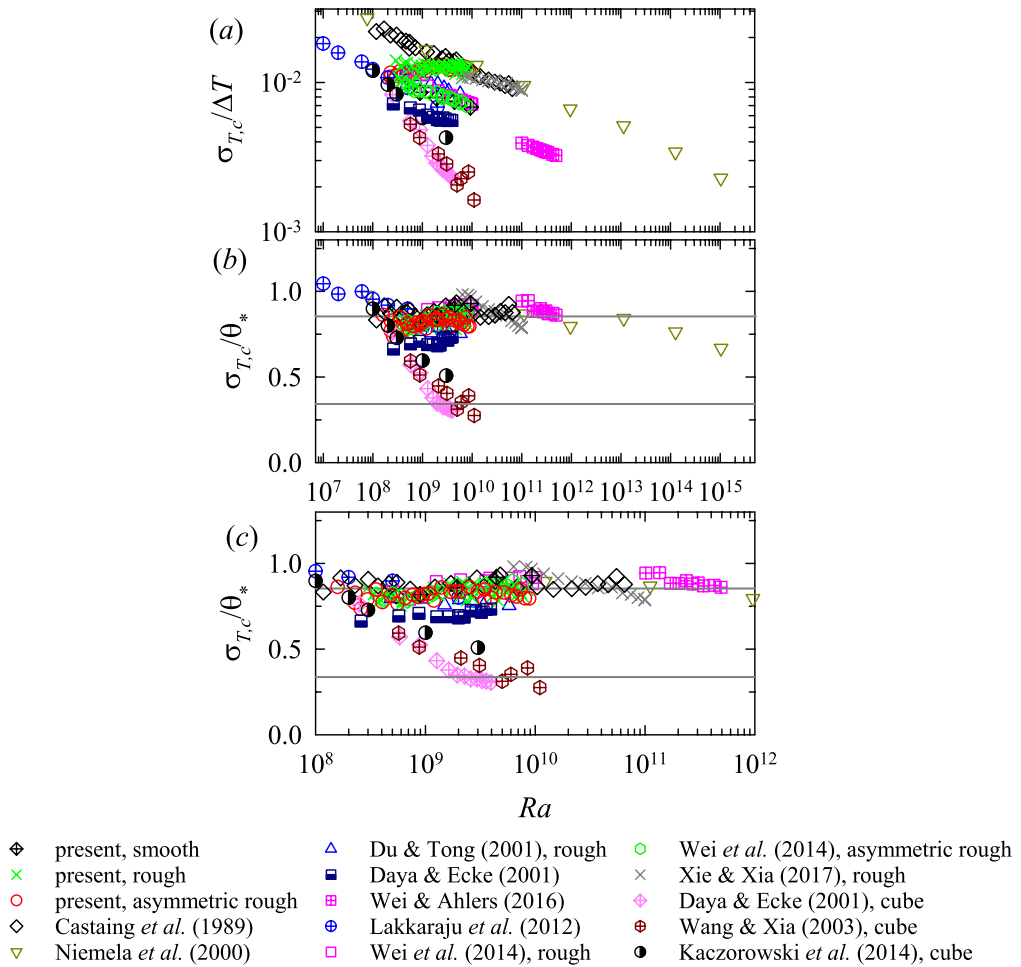}
      \caption{($a$) Normalised rms temperature at the cell centre $\sigma_{T,c}/\Delta T$ versus $Ra$. The legends with `rough' mean rough cells and those with `cube' means cubic cells. Details of the various data and the associated power laws are given in table \ref{Tab1}; ($b$) The same data as in ($a$) but with the vertical axis scaled by $\theta_*$. The upper and lower horizontal lines mark the averaged value of 0.85 in cylindrical cells and 0.34 in cubic cells, respectively. ($c$) A zoom-in of the region with $10^8\leq Ra\leq 10^{12}$. }
      \label{fig3}
    \end{figure}   
    
As $\sigma_T/\theta_*$ in the core region is a constant (figure \ref{fig1}$b$), we next focus on its value at the cell centre. Figure \ref{fig3}($a$) plots $\sigma_{T,c}/\Delta T$ versus $Ra$ measured in smooth cells and rough cells from present study. For comparison, $\sigma_{T,c}/\Delta T$ adopted from a number of sources are also shown. These data span $10^7\leq Ra\leq 10^{15}$ and $0.7\leq Pr\leq 23.34$, and were measured in smooth cells and rough cells. Details about their $Ra$, $Pr$, scaling property and cell type can be found in table \ref{Tab1}. 
  
When scaled with $\Delta T$, both the magnitude and the scaling property of $\sigma_{T,c}$ are seen to vary dramatically among different experiments (figure \ref{fig3}$a$). The data reveal that $\sigma_{T,c}/\Delta T$ generally decreases with $Pr$. Present measurements in rough cells even show a positive $\gamma$, i.e. $\sigma_{T,c}/\Delta T\sim Ra^{0.1}$. Figure \ref{fig3}($b$) plots the same data as those in figure \ref{fig3}($a$), but scaled with $\theta_*$ and figure \ref{fig3}($c$) shows a zoomed region with $10^8 \leq Ra\leq 10^{12}$. To obtain $\sigma_{T,c}/\theta_*$, we first took published data of $\sigma_{T,c}/\Delta T$ as a function of $Ra$ and $Pr$ from the cited references,  then calculated $\theta_*/\Delta T$ using equation \ref{equ1} with $Ra$, $Pr$ and $Nu$ that were measured together with $\sigma_{T,c}$. The ratio $(\sigma_{T,c}/\Delta T)/(\theta_*/\Delta T)$ is therefore $\sigma_{T,c}/\theta_*$.
 
We first look at data measured in cylindrical cells. It is seen that $\sigma_{T,c}/\theta_*$ from different experiments collapses around a straight line for $Ra>1\times10^8$ (in the so-called `hard turbulence’ regime), suggesting the existence of a universal constant. The upper solid line in figure \ref{fig3}($b$) marks the mean value, i.e. $\langle\sigma_{T,c}/\theta_*\rangle_{Ra,Pr}=0.85$. A mean value close to unity suggests that $\theta_*$ is indeed a representative temperature in the core region. This universal constant indicates that a common mechanism governs temperature fluctuation dynamics in the core region. 

For $\sigma_{T,c}/\theta_*$ measured in cubic cells, one sees that they decrease with $Ra$ for $Ra\leq 10^9$ and appear to reach another plateau around $\langle\sigma_{T,c}/\theta_*\rangle_{Ra,Pr}=0.34$ afterwards (the lower solid line in figure \ref{fig3}$b$). If this feature can be verified by data for higher $Ra$, then this means the above constant is cell geometry dependent. This dependence may be a result of the different azimuthal dynamics of the large-scale flow in the two geometries. A previous study suggests that the scaling of temperature fluctuations depend on cell geometry \citep{Daya2001PRL}. Now one sees that $\sigma_{T,c}/\theta_*$ reaches different plateaus in cells with different geometries. It can thus be used to characterise and quantify the level of turbulent fluctuations in different cell geometries. The apparent different behaviours in the bulk fluctuation remain to be explained. We note a potential application of the above result is in cases where it is difficult to measure $\sigma_T$ directly. In such situations, one can use the global quantities like $Ra$, $Nu$ and $Pr$ to obtain an estimate of the level of turbulent fluctuations in the core region.

       \begin{figure}
      \center
     \includegraphics[width=\textwidth]{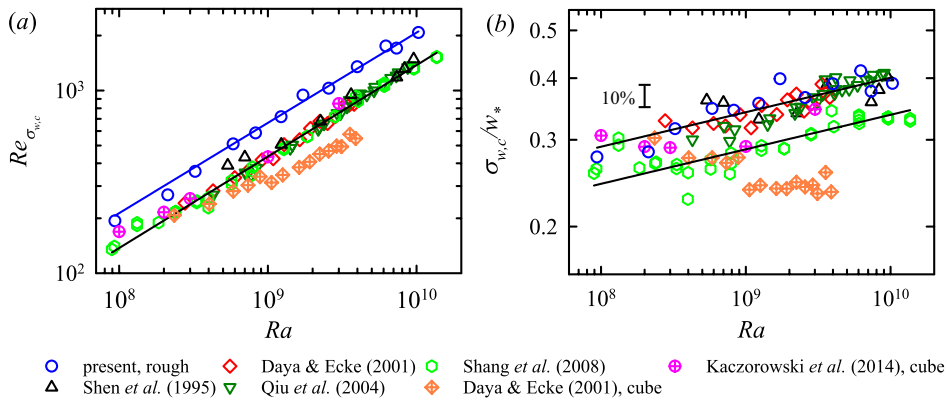}
      \caption{($a$) Measured Reynolds number $Re_{\sigma_{w,c}}$ in a rough cell. $Re_{\sigma_{w,c}}$ obtained in smooth cells from literatures are also shown. The solid lines are power law fits to the data yielding $Re_{\sigma_{w,c}}=0.021Ra^{0.50\pm0.01}$ (circles, rough cell) and $Re_{\sigma_{w,c}}=0.014Ra^{0.50\pm0.01}$ (smooth cell). ($b$) The $\sigma_{w,c}$ normalised by $w_*$ versus $Ra$. The two solid lines are power law fits to the data, i.e. $\sigma_{w,c}/w_*=0.08Ra^{0.07\pm0.02}$ (upper line) and $\sigma_{w,c}/w_*=0.07Ra^{0.07\pm0.02}$ (lower line).}
      \label{fig4}
    \end{figure}
    
\subsection{The vertical velocity fluctuation in the bulk}   

We study the velocity fluctuations in this section. Figure \ref{fig4}($a$) shows the Reynolds number $Re_{\sigma_{w,c}}$ based on the vertical velocity fluctuation $\sigma_{w,c}$ measured at the centre of the rough cell. For comparison, $Re_{\sigma_{w,c}}$ obtained in smooth cells from \citet{Shen1995PRL}, \citet{Daya2001PRL}, \citet{Qiu2004PoF} and \citet{Shang2008PRL} in cylinders and from \citet{Daya2001PRL} and \citet{Kaczorowski2014JFM} in cubes are also shown. The $Re_{w,c}$ in both smooth and rough cells with $\Gamma=1$ can be described by power laws, i.e. $Re_{\sigma_{w,c}}=0.021Ra^{0.50\pm0.01}$ (rough cell) and $Re_{\sigma_{w,c}}=0.014Ra^{0.50\pm0.01}$ (smooth cell). Note a transition is seen for $Ra\geq10^9$ in the data obtained in a cube by \cite{Daya2001PRL}, implying a geometry dependence of the velocity fluctuation. As these data are taken in a  cell with $\Gamma=0.7$, they are not used in the fitting. The velocity fluctuations in the rough cell are enhanced by 50\% when compared with those in smooth cells. This directly confirms that not only the temperature fluctuations are enhanced in a rough cell, but also the velocity fluctuations. 

To test if the convective velocity $w_*$ is a characteristic scale for velocity in the core region, we study $\sigma_{w,c}/w_*$. The $\sigma_{w,c}/w_*$ is obtained by that we first calculating $Re_{\sigma_{w,c}}$, and then $Re_{w_*}$ using equation \ref{equ1} with $Ra$, $Pr$ and $Nu$ that were measured simultaneously with $\sigma_{w,c}$. The ratio between $Re_{\sigma_{w,c}}$ and $Re_{w_*}$ is therefore $\sigma_{w,c}/w_*$. Figure \ref{fig4}($b$) plots $\sigma_{w,c}/w_*$ versus $Ra$. For the data from \citet{Daya2001PRL}, \citet{Qiu2004PoF} and \citet{Shang2008PRL}, there is no $Nu$ data available. To obtain $Re_{w*}$, we used the heat transport scaling relation $Nu=0.14Ra^{0.297}Pr^{-0.03}$, which was obtained in the $Ra$ range $2\times10^7\leq Ra\leq 3\times10^{10}$ and the $Pr$ range $4\leq Pr\leq 1350$ \citep*{Xia2002PRL}. For the rest of the data, $Re_{\sigma_{w,c}}$ and the corresponding $Nu$ were measured simultaneously. Interestingly, the data in rough cells and smooth cells collapse onto each other, suggesting that $\sigma_{w,c}/w_*$ exhibits universal behaviours that is independent of the plate topography. The scaled $\sigma_{w,c}$ shows a rather weak $Ra$ dependence, i.e. $\sigma_{w,c}/w_*\sim Ra^{0.07\pm0.02}$, as indicated by the solid lines in the figure. Note that the data from \cite{Shang2008PRL} are $\sim 14\%$ lower than the others. We currently do not understand this small difference. It may be due to the systematic error introduced when calculating $w_*$, which involves $Nu$ that was not measured simultaneously with $\sigma_{w,c}$.

        \begin{figure}
       \center
      \includegraphics[width=\textwidth]{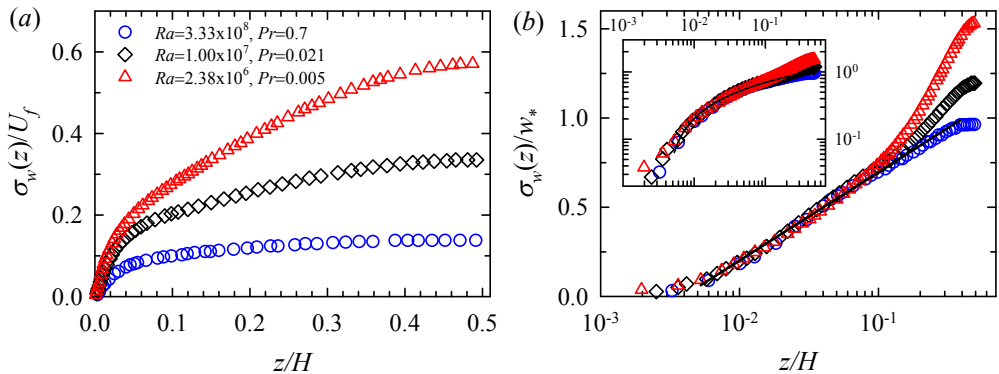}
      \caption{Profiles of vertical rms velocity $\sigma_w$. The vertical axes are scaled  by ($a$) the free fall velocity $U_f$ and ($b$) the convective velocity $w_*$. The solid line in ($b$) is logarithmic fit to the data in the range $4\times10^{-3}\leq z/H\leq 7\times10^{-2}$, yielding $\sigma_w/w_*=0.22 \ln(z/H)+1.22$. Inset of ($b$): $\sigma_w(z)/w_*$ in a log-log plot. The data are taken from \citet{Scheel2016JFM}.}
      \label{fig5}
    \end{figure}

Next, we examine the profiles of the vertical rms velocity $\sigma_w(z)$ in the mixing zone. Experimentally, obtaining velocity rms profiles requires long-time and multi-point measurement. Because of this, experimentally measured profiles of $\sigma_w$ are scarce. For this reason, we use profiles from numerical studies. Figure \ref{fig5}($a$) shows the horizontally averaged vertical rms velocity profiles $\sigma_w(z)$, which are adopted from \citet{Scheel2016JFM}. Both $Ra$ and $Pr$ for these profiles vary over a wide range. It is seen that, when normalised by the free-fall velocity $U_f$, the profiles do not collapse onto a single curve, suggesting that $U_f$ is not able to capture the essential physics here. Figure \ref{fig5}($b$) plots the same data set as those in figure \ref{fig5}($a$) but with the vertical axis scaled by $w_*$. The data for different $Ra$ and $Pr$ now collapse onto each other in the mixing zone, suggesting that $w_*$ is a proper velocity scale. The $\sigma_{w,c}/w_*$ in the range $4\times10^{-3}\leq z/H\leq 7\times10^{-2}$ can be fitted by a logarithmic function, i.e. $\sigma_w/w_*=0.22 \ln(z/H)+1.22$. This logarithmic dependence is in agreement with the theoretical prediction by \cite{Adrian1996}. It is also observed that the logarithmic region increases with $Ra$ and $Pr$. The inset of figure \ref{fig5}($b$) plots the same data as the main figure in log-log scale, showing that the data can not be fitted by a power law. Note $\sigma_{w,c}/w_*$ in the core region are larger than those shown in figure \ref{fig4}($b$) and they do not collapse onto each other. A possible reason may be that the data presented in figure \ref{fig4} were taken at a single point, i.e. the cell centre, but those in figure \ref{fig5} were averaged along a horizontal cross section which includes strong velocity fluctuations produced by thermal plumes carried with the large-scale circulation. 

\subsection{Implications for ultra-high-Rayleigh-number convection}

\begin{figure}
      \center
     \includegraphics[width=\textwidth]{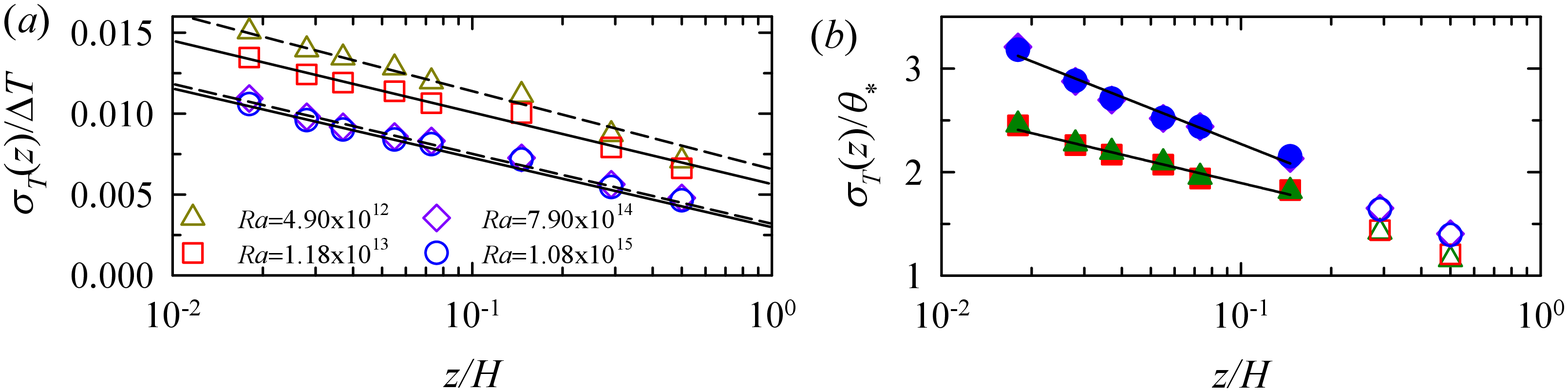}
      \caption{The rms temperature profiles measured in the ultra-high-$Ra$ convection. ($a$) Figure adapted from \cite{Ahlers2012PRL}. ($b$) The same set of data as in ($a$) with the vertical axis scaled by $\theta_*$. The solid lines in ($b$) are logarithmic fits to the data with solid symbols, i.e. $\sigma_T(z)/\theta_*=-0.49\ln(z/H)+1.15$ (upper line) and $\sigma_T(z)/\theta_*=-0.30\ln(z/H)+1.20$ (lower line). }
      \label{fig6}
    \end{figure}

We now study whether the convective temperature is applicable to the ultra-high-$Ra$ ($\geq10^{13}$) data and may possibly shed some lights on the turbulence in this regime. The data are taken from \citet{Ahlers2012PRL} with an adapted plot shown in figure \ref{fig6}($a$). Those data were measured in pressurised SF6 gas with $Ra$ reaching $1\times 10^{15}$ and $Pr \approx 0.8$. It is seen that the logarithmic functions fit $\sigma_T(z)/\Delta T$ nicely. But data for different $Ra$ do not collapse onto each other, and they seem to have similar decay rates (the pre-factor of the logarithmic term). The observed logarithmic dependence in the plume abundant region, i.e. near the sidewall, is also obtained by \cite{He2019PRL} recently. In figure \ref{fig6}($b$), we show that once $\sigma_T(z)$ is scaled by $\theta_*$, they fall into two groups. The solid lines in the figure are logarithmic fits to the data in the range $1.8\times10^{-2}\leq z/H\leq 1.46\times 10^{-1}$, yielding $\sigma_T(z)/\theta_*=-0.49\ln(z/H)+1.15$ for $Ra\geq 7.90\times10^{14}$ and $\sigma_T(z)/\theta_*=-0.30\ln(z/H)+1.20$ for $Ra\leq 1.18\times10^{13}$. The difference in the decay rates suggests that the bulk fluctuations undergo a transition. We note that \citet{He2012PRL} have stated that the system has reached the ultimate state of thermal convection for $Ra\geq 5\times10^{14}$. A systematic investigation on the $Ra$-dependence of $\theta_T(z)/\theta_*$ could provide more evidences on the existence of an internal flow state transition.

\section{Conclusions}\label{sec_con}

We have studied the temperature and the velocity fluctuations in the bulk of Rayleigh-B\'enard turbulence using new experimental data from the present study and experimental and numerical data from previous studies. We show that, when scaled by the convective temperature $\theta_*$, the rms temperature at the cell centre is a constant, i.e. $\sigma_{T,c}/\theta_*\approx 0.85$, over the Rayleigh number range of $10^{8}\leq Ra\leq 10^{15}$ and the Prandtl number range of $0.7\leq Pr \leq 23.34$, and is independent of the surface topographies of the top and bottom plates of the convection cell. A constant close to unity suggests that the convective temperature is a proper measure of the temperature fluctuation in the core region. The vertical rms velocity, on the other hand, shows a rather weak $Ra$-dependence, i.e. $\sigma_{w,c}/w_*\sim Ra^{0.07\pm0.02}$, which is also independent of the plate topography. In the mixing zone with a mean horizontal shear, the rms temperature profile $\sigma_T(z)/\theta_* $ obeys a power-law dependence on the vertical distance $z$ from the plate, and the vertical rms velocity profile $\sigma_w(z)/w_*$ obeys a logarithmic dependence on $z$. The study thus demonstrates that the typical scales for the temperature and the velocity are the convective temperature $\theta_*$ and the convective velocity $w_*$, respectively. The discovery of these universal aspects of fluctuations sheds new light on the bulk dynamics in convective turbulence. We further show that $\theta_*$ could be used to study internal flow state transitions in the ultra-high-$Ra$ turbulent convection. We note that, despite universal properties hold over a wide range of $Ra$ and $Pr$ and surface topographies, $\sigma_{T,c}/\theta_*$ is found to depend on the cell geometry, i.e. it is a constant of 0.34 in cubes and 0.85 in cylinders. The scaling exponent of the $\sigma_T/\theta_*$ profile also depends on the cell geometry. Finally, the present study focuses on turbulent fluctuations in convection cells with an aspect ratio around unity. It will be interesting to study if the universal behaviours observed here will also exist in cells with varying aspect ratios.

\section*{Acknowledgement}
We thank S.-D. Huang and Y.-H. He for discussions. This work was supported by a SUSTech Startup Fund and by the Hong Kong Research Grant Council under grant Nos. CUHK 14301115 and 14302317.

\end{document}